\documentclass[12pt]{article}
\usepackage{amssymb,amsmath,bm}
\usepackage{epsf}
\usepackage{epsfig}
\usepackage{graphicx}
\usepackage{psfrag}
\usepackage{cancel}
\usepackage{hyperref}
\usepackage{url}
\def\dfrac#1#2{{\displaystyle {#1 \over #2}}}

\def\simge{\mathrel{\rlap{\raise 0.511ex \hbox{$>$}}{\lower 0.511ex \hbox{$\sim$}}}}
\def\simle{\mathrel{\rlap{\raise 0.511ex \hbox{$<$}}{\lower 0.511ex \hbox{$\sim$}}}}
\def\slash#1{\setbox0=\hbox{$#1$}\dimen0=\wd0
     \setbox1=\hbox{/} \dimen1=\wd1 \ifdim\dimen0>\dimen1
     \rlap{\hbox to \dimen0{\hfil/\hfil}} #1                        \else
     \rlap{\hbox to \dimen1{\hfil$#1$\hfil}}
     /   \fi}

\newcommand{\lsim}{
\mathrel{\hbox{\rlap{\hbox{\lower4pt\hbox{$\sim$}}}\hbox{$<$}}}}

\newcommand{\gsim}{
\mathrel{\hbox{\rlap{\hbox{\lower4pt\hbox{$\sim$}}}\hbox{$>$}}}}

\allowdisplaybreaks[2]

\newcommand{\be}{\begin{equation}}
\newcommand{\ee}{\end{equation}}
\newcommand{\bea}{\begin{eqnarray}}
\newcommand{\eea}{\end{eqnarray}}
\newcommand{\nn}{\nonumber}
\newcommand{\bi}{\begin{itemize}}
\newcommand{\ei}{\end{itemize}}

\def\infn{a}
\def\infni{b}
\def\rmiii{c}
\def\rmi{d}

\begin{document}
\begin{titlepage}
{
\normalsize
\begin{center}
  \begin{Large}
    \textbf{\boldmath Power corrections to the CP-violation parameter $\varepsilon_K$\unboldmath} \\
  \end{Large}
\end{center}

\baselineskip 20pt plus 2pt minus 2pt
\begin{center}
  \textbf{
    M.\,Ciuchini$^{(\infn)}$,
    E.\,Franco$^{(\infni)}$,
    V.\,Lubicz$^{(\rmiii,\infn)}$,\\
    G.\,Martinelli$^{(\rmi,\infni)}$,
    L.\,Silvestrini$^{(\infni)}$,
    C.\,Tarantino$^{(\rmiii,\infn)}$
}\\
\end{center}

\begin{center}
  \begin{footnotesize}
    \noindent

$^{(\infn)}$ INFN, Sezione di Roma Tre,  Rome, Italy

$^{(\infni)}$ INFN, Sezione di Roma, Rome, Italy

$^{(\rmiii)}$ Dipartimento di Matematica e Fisica, Universit{\`a} Roma Tre, Rome, Italy

$^{(\rmi)}$ Dipartimento di Fisica, Universit{\`a} La Sapienza, Rome, Italy

  \end{footnotesize}
\end{center}

\begin{abstract}
We present the calculation of the short-distance  power corrections to the CP-violation parameter $\varepsilon_K$ coming from dimension-8 operators in the $\Delta S=2$ effective Hamiltonian. A first estimate of this contribution, obtained for large-$N_c$ and in the chiral limit, was provided in ref.\,\cite{Cata:2004ti}. Here we  evaluate and include  the $\mathcal{O}(m_K^2/m_c^2)$ and $\mathcal{O}(\Lambda_{QCD}^2/m_c^2)$  contributions that, a priori, could induce $\mathcal{O}(1)$ corrections to  previous estimates, as $m_K$ is numerically of order $\Lambda_{QCD}$. Our computation shows that there are several  dimension-8 operators besides the one considered before. Their effect on $\varepsilon_K$, however, accidentally cancels out to a  large extent, leaving the final correction at the level of 1\%.

\end{abstract}
}
\end{titlepage}

\section{Introduction}
The parameter $\varepsilon_K$, which describes indirect CP-violation in the $K^0 - \bar K^0$ system, represents one of the most interesting observables in Flavor Physics.  It plays an important role in the Unitarity Triangle Analysis  both within the Standard Model  and beyond,  see\,\cite{Ciuchini:2000de}-\cite{Alpigiani:2017lpj}, references therein and \cite{UTfit}.
As the phenomenon of $K^0 - \bar K^0$ mixing is a loop process further suppressed by the GIM mechanism, $\varepsilon_K$ turns out to be a powerful constraint on New Physics models, for which it is important to have experimental and theoretical uncertainties well under control.

On the one hand $\varepsilon_K$ is experimentally measured with a $0.4$\% accuracy, 
$\varepsilon_K^{exp}=(2.228 \pm 0.011) \cdot 10^{-3}$\,\cite{Zyla:2020zbs}, on the other hand, the theoretical accuracy of the SM prediction is approaching a few percent level, mainly thanks to the improvement of the lattice determination of the relevant bag parameter $B_K$\,\cite{Aoki:2019cca}.
Given the improved accuracy, the $\xi$ term and the deviation of the the phase $\phi_\varepsilon$ from $45^\circ$ appearing in the theoretical expression
\be
\varepsilon_K = e^{i \, \phi_\varepsilon} \sin \phi_\varepsilon \, \left( \frac{\mathrm{Im}M_{12}}{\Delta m_K} + \xi \right)\,,
\ee
are not negligible\,\cite{Buras:2008nn}.
Furthermore, the dominant long-distance contribution to $M_{12}$, due to the exchange of two pions, has been evaluated in\,\cite{Buras:2010pza}.
The inclusion of these three contributions gives a $6$\% reduction of the predicted central value of $\varepsilon_K$.
Another improvement producing a 2\% reduction of the central value is the inclusion of the available NNLO QCD corrections to the Wilson coefficients of the $\Delta S=2$ effective Hamiltonian\,\cite{Brod:2010mj,Brod:2011ty}. 
We have not included these corrections, however, since the relevant matrix element, $B_K$,  computed on the lattice  is matched to the $\overline{MS}$ coefficient at the NLO only. While the NNLO result for the Wilson coefficients is a crucial step to reduce the theoretical uncertainty of $\varepsilon_K$, at present there is no gain in evaluating the Wilson coefficient with NNLO accuracy when the accuracy of the matrix element is only at NLO: The overall uncertainty on $\varepsilon_K$ remains at the NLO level in any case. In this respect,  the perturbative calculation of the NNLO matching of $B_K$  from the non-perturbative  RI-MOM/SMOM schemes used in lattice calculations to the $\overline{MS}$ scheme would be welcome. As discussed in ref.~\cite{Brod:2019rzc}, the NNLO QCD correction for the charm contribution is actually not needed.

Given the increasing precision of the theoretical ingredients entering $\varepsilon_K$, it is becoming important to include all terms expected to contribute to the theoretical evaluation of $\varepsilon_K$ at the percent level.
In this paper,  we focus on the power corrections due to the finite value of  the charm quark mass, denoted as $\delta_{m_c}$, coming from dimension-8 operators in the effective Hamiltonian. 
The naive dimensional estimate of $\delta_{m_c}$ is of $\mathcal{O}(m_K^2/m_c^2)\sim 15$\%.   Its size, however,  is reduced to about $2$-$3$\%, because the dominant top quark contribution to  $\varepsilon_K$ is unaffected by these corrections.
A first estimate of $\delta_{m_c}$, obtained for large-$N_c$ and in the $SU(3)$ chiral limit, has been presented in\,\cite{Cata:2004ti}. In these limits, there is only one dimension-8 operator contributing to $\delta_{m_c}$\,\cite{Pivovarov:1991bh}. Numerically, ref.\,\cite{Cata:2004ti} finds $\delta_{m_c}\sim 0.5 - 1$\% and it has been neglected in phenomenological analyses up to now.
 $\mathcal{O}(1)$ corrections to this estimate may, however,  be expected, because $m_K$ is numerically of order $\Lambda_{QCD}$. 
In addition, in the large-$N_c$ limit, an operator non vanishing in the chiral limit is also discarded. 
Indeed the full calculation, presented in this paper, shows that there are several new operators contributing to $\delta_{m_c}$.   We find,  however,  that their effect on $\delta_{m_c}$ accidentally cancels out to a large extent, leaving the final correction at the level of 1\%.

The paper is organized as follows. In section \ref{sec:outline} we discuss the $\Delta S=2$ effective Hamiltonian relevant for the calculation,  including dimension-8 operators. In section \ref{sec:matching} we describe the matching procedure followed to compute the Wilson coefficients of the operators entering $\delta_{m_c}$. In section \ref{sec:via} we present the evaluation of the matrix elements of the relevant operators in the vacuum insertion approximation (VIA), on which we rely lacking more precise determinations of the matrix elements  at  the leading order in chiral perturbation theory.   In section \ref{sec:possible} we propose a new approach to compute non-local contributions on the lattice, while in section \ref{sec:prospects} we estimate $\delta_{m_c}$ and comment on its impact on the theoretical prediction of $\varepsilon_K$. Some more technical details, such as the relations that have been used in the computation, based on equations of motion and Fierz transformations, are collected in three appendices.

\section{Outline of the calculation}
\label{sec:outline}
\begin{figure}[t]
\centering
\includegraphics[width=15.0cm]{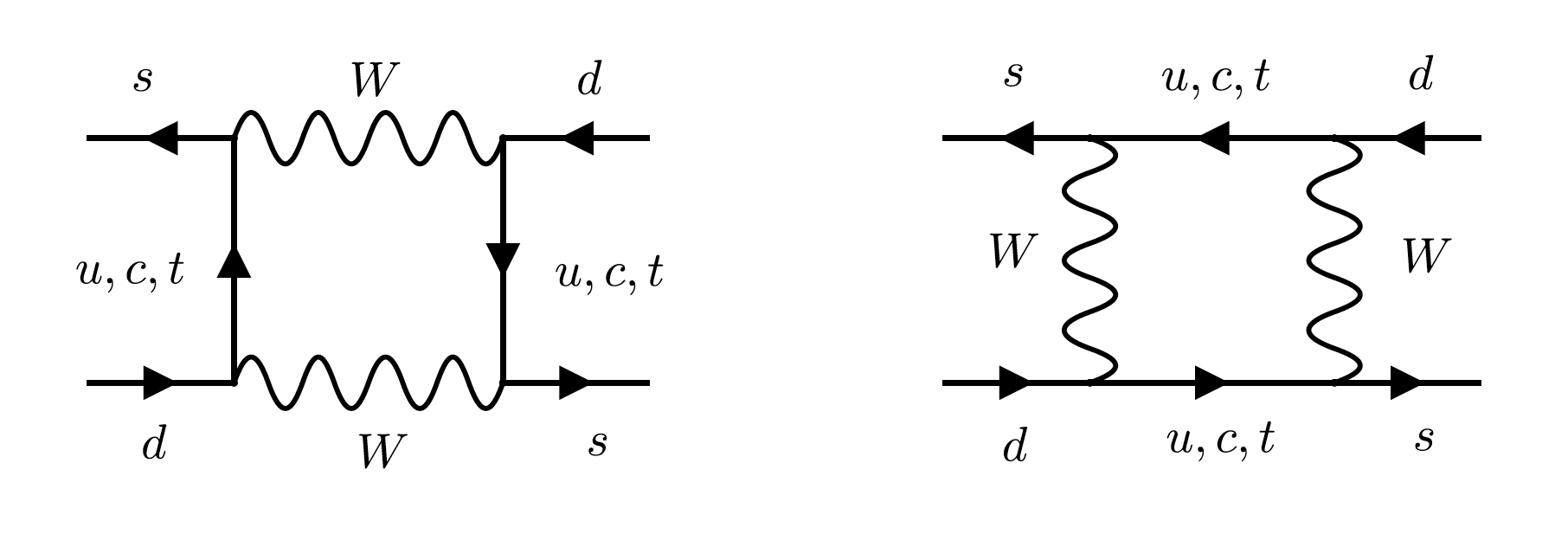}
\caption{\it Box diagrams describing $K^0 - \bar K^0$ mixing}
\label{fig:boxes}
\end{figure}

In this work we calculate the corrections of $\mathcal{O}(m_K^2/m_c^2)$  and $\mathcal{O}(\Lambda_{QCD}^2/m_c^2)$,   relative to  the dominant terms,  to the $\Delta S=2$ effective Hamiltonian describing $\varepsilon_K$. For simplicity, in the following we will denote the ensemble  of  these $\mathcal{O}(1/m_c^2)$ corrections as  $\delta_{m_c}$.

The imaginary part  of the $\Delta S=2$ effective Hamiltonian at the leading order in the Operator Product Expansion (OPE) and at zero order  in the strong interactions, has the well known expression
\bea
H^{ \mathrm{eff}}_{LT} &=& \frac{G_F^2}{16\, \pi^2} \, M_W^2 \, \left[ 
 \mathrm{Im} \left(\lambda_c^2\right) \,  S_0(x_c)  +  \mathrm{Im} \left(\lambda_t^2\right) \,  S_0(x_t) +\right. \nn \\
&& \left.  + 2\,  \mathrm{Im}\left(\lambda_c \lambda_t\right) \, S_0(x_c,x_t)\right] \, (\bar s \gamma^\mu_L d)\,(\bar s \gamma^\mu_L d)\,,
\label{eq:leadS0}
\eea
where  the subscript $LT$ means dominant  (leading)  term in the $1/m_c$ expansion; $\lambda_i = V_{id} V^*_{is}$ and the short-distance Inami-Lim  functions $S_0$'s are computed by matching full and effective amplitudes for external states of four quarks with zero momentum\,\cite{Inami:1980fz}. The full amplitude is represented by the box diagrams shown in fig.\,\ref{fig:boxes}.

In eq.\,(\ref{eq:leadS0}),  the CKM factor $\lambda_u$ has been eliminated by using the unitarity relation $\lambda_u+\lambda_c+\lambda_t=0$ and, thus, the Inami-Lim  functions  are linear combinations of the functions ($D_{ij}$) corresponding to box diagrams with internal $i$ and $j$ quarks, namely
\bea
S_0(x_c)&=&D_{uu}+D_{cc}-2 D_{uc}\,,\nn\\
S_0(x_t)&=&D_{uu}+D_{tt}-2 D_{ut}\,,\nn\\
S_0(x_c,x_t)&=&D_{uu}-D_{uc}-D_{ut}+D_{ct}\,.
\eea

An equivalent way to express the $\Delta S=2$ effective Hamiltonian comes from the fact that in the standard convention  for the CKM matrix\,\cite{Zyla:2020zbs} $\lambda_u$ is real, so that $\mathrm{Im}(\lambda_u^2)=\mathrm{Im}(\lambda_c^2+\lambda_t^2+2\lambda_c \lambda_t) =0$. By using this relation in eq.\,(\ref{eq:leadS0}), one can write
\bea
H^{ \mathrm{eff}}_{LT} 
&=& \frac{G_F^2}{16\, \pi^2} \, M_W^2 \, \left[ 
 \mathrm{Im}\left(\lambda_c^2\right) \, \left(  S_0(x_c) -  S_0(x_c,x_t) \right) + \right. \nn \\
&& \left. +  \mathrm{Im}\left(\lambda_t^2\right) \, \left(  S_0(x_t) -  S_0(x_c,x_t) \right) \right] \, (\bar s \gamma^\mu_L d)\,(\bar s \gamma^\mu_L d) \,.
\label{eq:lead2CKM}
\eea

At the leading order  in the $1/m_c$ expansion  both the $\mathrm{Im}(\lambda_c^2)$ and $\mathrm{Im}(\lambda_t^2)$ contributions are relevant. The latter, in fact, which is CKM suppressed, is enhanced by the top quark mass, as $S_0(x_t)$ grows as  $x_t=m_t^2/M_W^2$  in the large $x_t$ limit, while $S_0(x_c)$ and $S_0(x_c,x_t)$ vanish as   $x_c=m_c^2/M_W^2$ in the small $x_c$ limit.
For the subleading contributions due to dimension-8 operators, instead, only the $\mathrm{Im}(\lambda_c^2)$ term has some relevance. The $\mathrm{Im}(\lambda_c^2)$ subleading contribution, indeed, represents a correction of $\mathcal{O}(1/m_c^2)$ to the leading term, while in the $\mathrm{Im}(\lambda_t^2)$ term the relative correction is of $\mathcal{O}(1/m_t^2)$ and can  safely be neglected.

The calculation of the $\delta_{m_c}$  is  based on the determination of the Wilson coefficients of dimension-8 operators from matching conditions.
To this aim, box diagrams have to be calculated with non vanishing  external quark momenta. The $\mathcal{O}(p^2)$ contribution to the full amplitudes is then computed taking arbitrary external momenta and performing derivatives with respect to the external momenta.    Since we  are evaluating  small  $\mathcal{O}(1/m_c^2)$ corrections  to $\varepsilon_K$, we will  use the effective Hamiltonian expanded at zeroth order in the strong interactions. 
 At $\mathcal{O}(\alpha_s)$, apart from  the contribution of dimension-8 operators,   other  contributions cannot  be  written in terms  of  local operators,  as discussed in the following.

We consider from the beginning only  the case of $p^2  < m_c^2$.
The matching condition   schematically reads
\be
\mathcal{A}^{full}(\Delta S=2) = \mathcal{A}(\Delta S=1 \otimes \Delta S=1) + \sum_i C^W_i \langle O_i \rangle\,, \label{eq:mach}
\ee
where the first term on the r.h.s.  represents the long distance contribution to the amplitude $\mathcal{A}$,  coming from the double insertion of the  effective 
$\Delta S=1$ Hamiltonian,  whereas the last term is the contribution that can be expressed as linear combination of local operators multiplied by suitable Wilson coefficients $C^W_i$.   
By assuming the charm quark mass to be {\it large}, that is by neglecting terms of  $\mathcal{O}(1/m_c^2)$,  the first term  on the r.h.s. of eq.\,(\ref{eq:mach}) is absent and the second, in the Standard Model,  is simply given by the operator 
\be \bar{s}\gamma^{\mu}_L d\ \bar{s}\gamma^{\mu}_Ld \, , \ee
see eq.\,(\ref{eq:lead2CKM}).
\begin{figure}[t]
\centering
\includegraphics[width=16.0cm]{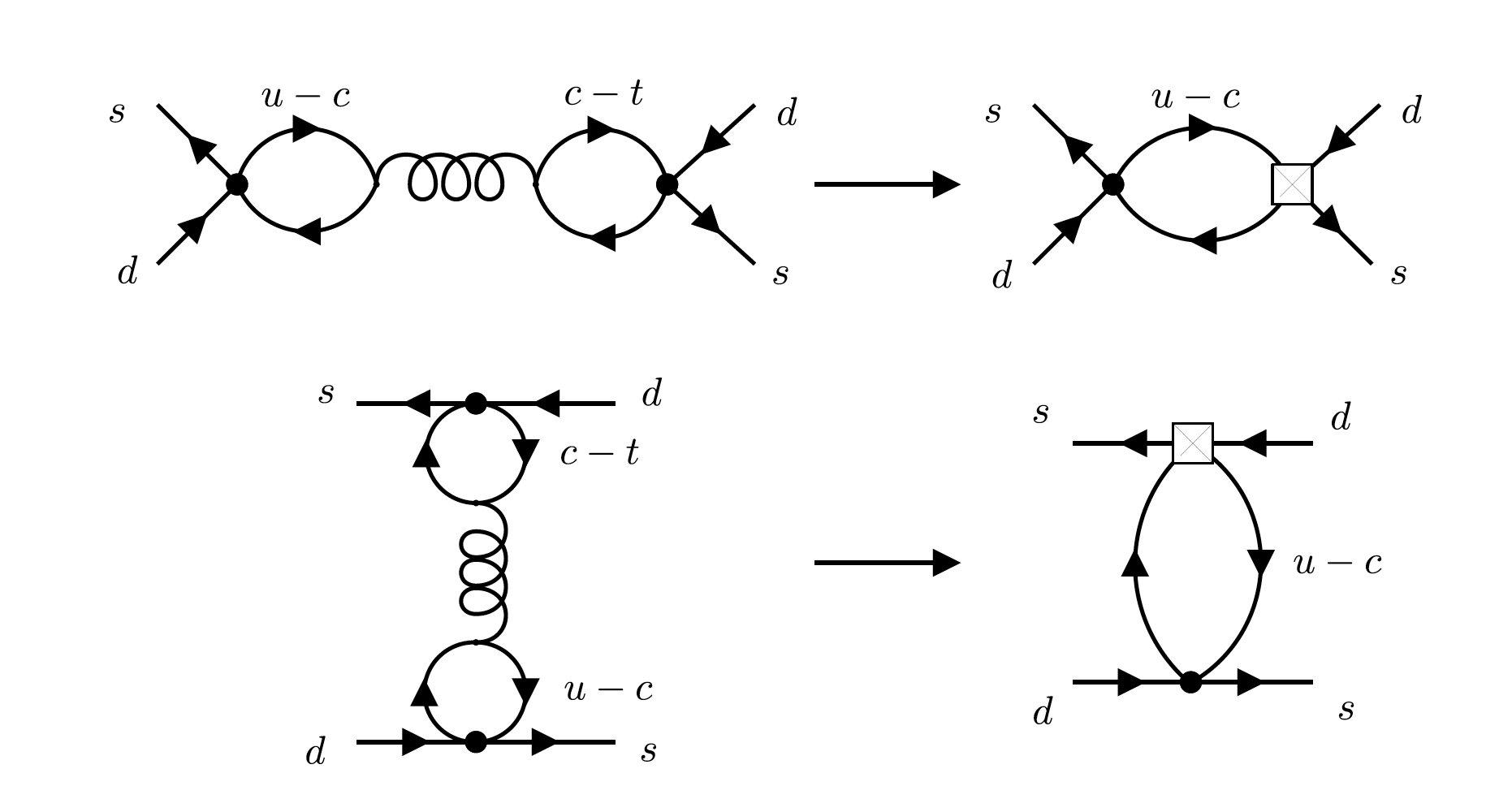}
\caption{\it On the left side the $K^0 - \bar K^0$   double penguin  diagrams, on the  right side   their reinterpretation in terms of  connected  diagrams.  The box represents the insertion of the penguin operator defined in eq.\,(\ref{eq:QGIM}).}
\label{fig:penaprox}
\end{figure}

When we include the  $\mathcal{O}(1/m_c^2)$ corrections,  in addition to the  presence of local dimension-8  operators,  we find a non-local, non-perturbative,  non-box  amplitude $\mathcal{A}(\Delta S=1 \otimes \Delta S=1) $ given by  double penguin diagrams  of  $\mathcal{O}(\alpha_s \, p^2/m_c^2)$   shown  in the left  panels of fig.\,\ref{fig:penaprox}. 
  The black dots in these  figures  represent the double insertion of the $\Delta S=1$ Hamiltonian.   
  Using the unitarity of the CKM matrix one may reduce all the double penguin diagrams to the GIM-like combination  of penguins, namely up minus charm and charm minus top, shown in the figure. In  general,  since the typical scale of these diagrams is of  $\mathcal{O}(p^2) \sim \mathcal{O}(\Lambda_{QCD}^2)$,  and the intermediate states propagating in the $s$ or $t$ channels are light,  perturbation theory  is of  no  use  and $\mathcal{A}(\Delta S=1 \otimes \Delta S=1) $ must be evaluated  non perturbatively.  This is to be contrasted  with the case of the  box diagrams  where the non-perturbative part is encoded in the matrix elements of local operators.  The  contribution to the mixing amplitude  of the double penguin diagrams of  figs.\,\ref{fig:penaprox}  corresponds to the long distance corrections to  $\varepsilon_K$ estimated in ref.\,\cite{Buras:2010pza}.   Note that this kind of  diagrams  are  also extremely difficult to compute from first principles,  in a lattice QCD simulation, even in the simpler case of  the $K_L$-$K_S$ mass difference  where only the operators  $Q_1$ and $Q_2$ of the $\Delta S=1$ Hamiltonian are considered.  
  Although there is no theoretical issue in the lattice calculation, the   difficulty is the standard one of achieving sufficient precision when evaluating disconnected diagrams. For more details   see for example  the discussion on the {\it disconnected} diagrams Type 3 and Type  4   of  refs.\,\cite{Christ:2014qwa,Bai:2018mdv},   where the reader can also find the definition of  the operators $Q_1$ and $Q_2$.  For a direct calculation of  $\varepsilon_K$  
  see ref.\,\cite{Christ:2015phf}.  In the present paper,  for  our final numerical estimate of the  correction at $\mathcal{O}(1/m_c^2)$ to $\varepsilon_K$ we will indeed  use the estimate  of $\mathcal{A}(\Delta S=1 \otimes \Delta S=1) $ by  ref.\,\cite{Buras:2010pza}.   Some further discussion about this non-local contribution will  be given   in section\,\ref{sec:possible}.
The  diagrams  in the right panels of  fig.\,\ref{fig:penaprox} will be discussed in section\,\ref{sec:prospects}.
    
\section{Matching}
\label{sec:matching}
 The Wilson coefficients that we want to determine can be derived by imposing  the following matching conditions  on amplitudes with  four quarks and  four quarks plus a gluon  external states respectively,
\be
\mathcal{A}^{full}_{4q}=\mathcal{A}^{ \mathrm{eff}}_{4q}\,,\qquad\qquad \mathcal{A}^{full}_{4q1g}=\mathcal{A}^{ \mathrm{eff}}_{4q1g}\,, 
\label{eq:match}
\ee
for some choice of external particle momenta  $p_i \ll m_c$,  where ``full" and ``effective" denote the amplitudes in the full (Standard Model) and effective theories respectively.

The full amplitudes at  $\mathcal{O}(\alpha^0_s)$ are calculated from the box diagrams of fig.\,\ref{fig:boxes}, with W propagators expanded at the lowest order in the large W mass, i.e. approximated by 4-fermion contact interactions. By choosing for the matching the color indices of the external quark and anti-quark states in a proper way, one can actually select the contribution of only one of the two box diagrams. For the four quark amplitude, for instance, we choose equal color indices for the quark and anti-quark in the initial state and similarly for the final state, but  different between the initial and final states, which corresponds to selecting the box diagram in fig.\,\ref{fig:boxes} (left panel). In this case, the amplitude has the form
\be
\bar s^a(p_2) \Gamma^i d^a(p_1) \bar s^b(p_3) \Gamma^j d^b(p_4)
\ee
where $p_1,p_2$ and $p_3,p_4$  are initial and final momenta respectively and $a$ and $b$ are different color indices. The calculation of the box diagram with the four quarks in the external states then leads to the expression
\be
\label{eq:Afulltot}  \mathcal{A}^{full}_{4q}=-i\,\frac{G_F^2}{16\, \pi^2} \, \mathrm{Im}(\lambda_c^2)
\,\left[\mathcal{A}^{full}_{4q}|_{\gamma^\mu\gamma^\mu}+\mathcal{A}^{full}_{4q}|_{\slash{q}\, \slash{q}}\right]\, ,
\ee
where, by omitting for brevity the external spinors, eliminating  the combination $p_4-p_2$ in  favour of $q=p_1-p_3$ and using the equations of motion,
\bea
&&\mathcal{A}^{full}_{4q}|_{\gamma^\mu\gamma^\mu}=\, \alpha_1\,(p_1-p_3)^2 \gamma^\mu_L \otimes \gamma^\mu_L=
\,\alpha_1\,(m_d^2+m_s^2-2p_1\cdot p_3) \gamma^\mu_L\otimes \gamma^\mu_L\,, \nn \\
&&\mathcal{A}^{full}_{4q}|_{\slash{q}\, \slash{q}}= \, \alpha_2\, (\slash{p_1}-\slash{p_3})_L \otimes (\slash{p_1}-\slash{p_3})_L\nn\\
&=&\, \alpha_2\, \left[ m_d\,R \otimes \slash{p}_{1\,L} -m_d m_s\,R \otimes L-\slash{p}_{3\,L}\otimes \slash{p}_{1\,L}+m_s\,\slash{p}_{3\,L} \otimes L \right]\,,
\eea
with
\be
\alpha_1=-\frac{5}{9}\,,\qquad\qquad \alpha_2=2\,\alpha_1=-\frac{10}{9}\,.
\ee
Throughout this paper we use the notation
\be
R \equiv 1 + \gamma^5 \qquad , \qquad L \equiv 1 - \gamma^5 \ .
\ee

At $\mathcal{O}(\alpha^1_s)$,  the amplitude for the four quarks and a gluon includes the contributions of a gluon emitted either from an internal line or from an external leg.
The expression for this amplitude is quite long and tedious, containing about 50 different Dirac structures. For this reason   it is not reported here but given in Appendix  C.
Having the full amplitudes at hand,  we then need the effective amplitudes in order to impose the matching conditions of eq.\,(\ref{eq:match}).
The computation of the effective amplitudes is based on the effective Hamiltonian
\be
H^{ \mathrm{eff}}_{1/m_c^2} = \frac{G_F^2}{16\, \pi^2} \, \mathrm{Im}(\lambda_c^2) \,\sum_i C_i \, O_i \,,
\label{eq:Heff1mc2}
\ee
where the $C^W_i= G_F^2/(16\, \pi^2) \, \mathrm{Im}(\lambda_c^2)\,C_i$ are the Wilson coefficients that we want to determine and the $O_i$ are the local  operators in the expansion\,\cite{Cata:2003mn}, namely
\bea\label{eq:operators} &&O_1=\bar{s}\gamma^{\nu}_L{\cal{D}}_{\mu} d\ \bar{s}\gamma^{\nu}_L{\cal{D}}^{\mu}d + 
\bar{s} \overleftarrow{{\cal{D}}_{\mu}}\gamma^{\nu}_Ld \ \bar{s}\overleftarrow{{\cal{D}}^{\mu}}\gamma^{\nu}_L d\ ,\nn \\
&&O_2=
\bar{s}\gamma^{\nu}_L{\cal{D}}_{\mu}d\ {\bar{s}}\gamma^{\mu}_L{\cal{D}}^{\nu}d  +
\bar{s}\overleftarrow{{\cal{D}}_{\mu}}\gamma^{\nu}_L d \ \bar{s}\overleftarrow{{\cal{D}}_{\nu}}\gamma^{\mu}_Ld  \ ,\nn \\
&&O_3=\bar{s}\overleftarrow{{\cal{D}}_{\mu}}\gamma^{\nu}_L d \ \bar{s}\gamma^{\nu}_L{\cal{D}}^{\mu}d \ ,\nn \\
&&O_4=g_s\, \bar{s}\gamma^{\mu}_L\,\tilde{G}_{\mu\nu} d \ \bar{s}\gamma^{\nu}_L d  \, \\
&&O_5=i\,m_s\,\bar{s}\,L\, {\cal{D}}_{\mu} d\ \bar{s}\gamma^{\mu}_L d -  
i\,m_d\,\bar{s}\overleftarrow{\cal{D}}_{\mu}\,R d\ \bar{s}\gamma^{\mu}_L d \ ,\nn \\
&&O_{6}=(m_s^2 + m_d^2) \,\bar{s} \gamma^{\mu}_L d\ \bar{s}\gamma^{\mu}_L d \ , \nn \\
&&O_{7}=m_s^2\,\bar{s} L d\ \bar{s}L d + m_d^2\,\bar{s} R d\ \bar{s} R d \nn \ , \nn \\
&&O_{8}=m_s\, m_d \,\bar{s} L d\ \bar{s} R d \nn \ ,
\label{eq:pivovarov}
\eea
with ${\cal{D}}_{\mu}=\partial_\mu-i g_s G^A_\mu t^A$,   $\tilde{G}^A_{\mu\nu} = 1/2\, \epsilon_{\mu \nu \rho \sigma}\, G^{ \rho \sigma\,A}$,  $G_{ \rho \sigma} = G_{ \rho \sigma}^A t^A=(\partial_\rho G^A_\sigma-\partial_\sigma G^A_\rho+ g_s\, f^{ABC} G^B_\rho G^C_\sigma) \,t^A$, implying a sum over repeated color indices, and ${\rm Tr}( t^A t^B) = 1/2\, \delta^{A B}$.

The effective amplitude for a state of four quarks has then the form
\be
\label{eq:Aeff4q}
\mathcal{A}^{ \mathrm{eff}}_{4q} = 
-\, i \frac{G_F^2}{16\, \pi^2} \, \mathrm{Im}(\lambda_c^2) \, \sum_i C_i \, \langle O_i\rangle_{4q} \,,
\ee
where the matrix elements for the various operators in eq.\,(\ref{eq:pivovarov}) read
\bea
&&\langle O_1 \rangle_{4q}=2\,(p_1\cdot p_4+p_2 \cdot p_3)\,\gamma^\mu_L \otimes \gamma^\mu_L=2\,(m_d^2-m_s^2+2\, p_2\cdot p_3)\,\gamma^\mu_L \otimes \gamma^\mu_L \,, \nn \\
&&\langle O_2 \rangle_{4q}=2\,(p_2^\mu p_3^\nu+p_1^\mu p_4^\nu)\,\gamma^\nu_L \otimes \gamma^\mu_L=\nn\\
&&\qquad\quad=2\,(-2\,\slash{p}_{3\,L} \otimes \slash{p}_{1\,L}+m_s\,\slash{p}_{3\,L} \otimes L-m_d\,\slash{p}_{3\,L} \otimes R-m_s\,L \otimes \slash{p}_{1\,L} + m_d\,R \otimes \slash{p}_{1\,L})\,, \nn \\
&&\langle O_3 \rangle_{4q}=(p_2 \cdot p_4+p_1 \cdot p_3)\,\gamma^\mu_L \otimes \gamma^\mu_L=2\,p_1 \cdot p_3\,\gamma^\mu_L \otimes \gamma^\mu_L \,, \nn \\
&&\langle O_4 \rangle_{4q}= 0\,,\\
&&\langle O_5 \rangle_{4q}=m_s^2\, L \otimes L +m_d^2\, R \otimes R - 2\,m_s\,m_d R \otimes L \nn\\
&& \qquad\qquad + m_s\,\slash{p}_{3\,L} \otimes L + m_d\,\slash{p}_{3\,L} \otimes R + m_s\,L \otimes \slash{p}_{1\,L} + m_d\,R \otimes \slash{p}_{1\,L}\,, \nn \\
&&\langle O_6 \rangle_{4q}=2\, (m_s^2+m_d^2) \,\gamma^\mu_L \otimes \gamma^\mu_L \,, \nn \\
&&\langle O_7 \rangle_{4q}=2\,m_s^2\,L \otimes L + 2\,m_d^2\,R \otimes R\,, \nn \\
&&\langle O_8 \rangle_{4q}= m_s\, m_d \,(L \otimes R + R \otimes L) \,. \nn
\eea
For the state of four quarks plus a gluon, the expressions of the matrix elements are quite long, see Appendix C,  and here we write the result only for the operator $O_4$
\be
\langle O_{4} \rangle_{4q1g}=g_s\,\epsilon^{\mu \nu \rho \sigma }(i\,q_\rho)\,\varepsilon_\sigma^A\,\left[ \gamma^\mu_L\,t^A \otimes \gamma^\nu_L - \gamma^\mu_L \otimes \gamma^\nu_L\,t^A \right]\,.
\ee
The operator $O_4$, at variance with the other seven operators, has vanishing matrix element on the four quark state. As a consequence, the determination of the Wilson coefficient $C_4$ requires the matching between the full and the effective amplitude on a state with a gluon.

From the matching conditions in eq.\,(\ref{eq:match}) we find
\bea
&&C_1=0\,,\qquad C_2=-\frac{5}{18}\,,\qquad C_3=\frac{5}{9}\,,\qquad C_{4}=- \frac{13}{9}\,, \nn\\
&&C_5=-\frac{5}{9}\,,\qquad C_6=-\frac{5}{18}\,,\qquad C_7=\frac{5}{18}\,,\qquad C_8=0 \,.
\eea

\section{Matrix elements in the VIA}
\label{sec:via}
In this section we evaluate the matrix elements of the relevant four fermion operators in the VIA which, lacking a more precise quantitative determination, is  expected to provide a reasonable estimate of the matrix elements. In this approximation the four fermion operators are   expressed in terms of  factored bilinear operators, the matrix elements of which, in this paper,    are   evaluated by  using the chiral effective theory, at the leading order. Given the level  of accuracy at which we are working  we will  consider neither the  QCD  corrections to the Wilson coefficients of the dimension-8 operators nor the usual  problems  related to the presence of renormalons in matching  the original theory to its OPE at the subleading order\,\cite{Martinelli:1996pk}.

In the chiral effective theory, one  starts by defining the meson fields and the explicit symmetry breaking term,
\be 
U = e^{2i \Pi/f } \qquad , \qquad  \chi =  2 B M
\ee
where $f$ is the pseudoscalar  decay constant defined such that $f_\pi\sim 130$~MeV, $M = {\rm diag}(\, m_u, \, m_d, \, m_s \, )$ and $\Pi$ is the pseudoscalar meson operator,
\be
\Pi =
\left(
\begin{array}{ccc}
 \eta/\sqrt{6} + \pi^0/\sqrt{2}  &   \pi^+  &   K^+ \\
 \pi^-  &   \eta/\sqrt{6} - \pi^0/\sqrt{2}  &   K^0 \\
 K^-  &   \bar K^0  &   -\sqrt{2/3}\, \eta
 \end{array}
\right) \ .
\ee
Under $SU(3)_L \otimes SU(3)_R$ chiral transformations, the fields $U$ and $\chi$ transform as $U \to R\, U\, L^\dagger$ and $\chi \to R\, \chi\, L^\dagger$.

Next we provide the expressions of the relevant  two-quark operators  in terms of the chiral fields, at the leading order in the chiral expansion. We start by considering the dimension-3 operators:
\bea 
\label{eq:dim3a}
&& \bar q_i \,  L \, q_j = - \frac{f^2}{2}B [\, U^\dagger \,]_{ji} \, , \\
\label{eq:dim3b}
&& \bar q_i \,  R \, q_j = - \frac{f^2}{2}B [\, U \,]_{ji} \, , \\
\label{eq:dim3c}
&& \bar q_i \, \gamma^\mu_L \, q_j = i\, \frac{f^2}{2} [\, (\partial^\mu U^\dagger) \, U\,]_{ji}  \, , \\
\label{eq:dim3d}
&& \bar q_i \, \gamma^\mu_R \, q_j = i\, \frac{f^2}{2} [\, (\partial^\mu U) \, U^\dagger \,]_{ji}  \, , \\
\label{eq:dim3e}
&& \bar q_i \, \sigma^{\mu\nu}_L \, q_j = - i\, \alpha_T \, \frac{f^2}{2} 
     [\, (\partial^\mu U^\dagger) \, (\partial^\nu U)\, U^\dagger\, - 
      \, (\partial^\nu U^\dagger) \, (\partial^\mu U)\, U^\dagger\, ]_{ji}  \, , \\
\label{eq:dim3f}
&& \bar q_i \, \sigma^{\mu\nu}_R \, q_j = - i\, \alpha_T \, \frac{f^2}{2} 
     [\, (\partial^\mu U) \, (\partial^\nu U^\dagger)\, U\, - 
      \, (\partial^\nu U) \, (\partial^\mu U^\dagger)\, U\, ]_{ji}  \ .
\eea
In the chiral expression of the tensor operators the low energy constant $\alpha_T$ is given by $\alpha_T=B_T/(2\, m_K)$ where $B_T$ is a dimensionless parameter of ${\cal O}(1)$\,\cite{Cata:2003mn}. Note, however, that these operators do not contribute in the VIA, since their matrix elements between the vacuum and a kaon state vanish.

We now list the dimension-4 operators\,\cite{BIS} at the lowest order in the chiral expansion:
\bea
\label{eq:buch1}
&& \bar q_i \,  \overleftarrow{D}^\alpha  \gamma^\mu_L \, q_j = 
i \, \frac{f^2}{4} \left[\, (\partial^\alpha \partial^\mu U^\dagger) \, U\, - 
\frac{1}{8} g_{\alpha\mu} (\chi^\dagger U + U^\dagger \chi)\right]_{ji} +\dots \, , \\
\label{eq:buch2}
&& \bar q_i \,   \gamma^\mu_L \, \overrightarrow{D}^\alpha q_j = 
i \, \frac{f^2}{4} \left[\, (\partial^\alpha \partial^\mu U^\dagger) \, U\, + 
\frac{1}{8} g_{\alpha\mu} (\chi^\dagger U + U^\dagger \chi)\right]_{ji}+\dots  \, ,
\eea
where on the r.h.s.  of  eqs.\,(\ref{eq:buch1}) and (\ref{eq:buch2}) we only show explicitly  the terms which contribute in the factorized case. Eq.\,(\ref{eq:buch2}) can be derived from (\ref{eq:buch1}) using the identity
\bea 
\partial^\alpha \left( \bar q_i \,   \gamma^\mu_L \,  q_j\right)  = q_i \,  \overleftarrow{D}^\alpha  \gamma^\mu_L \, q_j +
\bar q_i \,   \gamma^\mu_L \, \overrightarrow{D}^\alpha q_j \ .   
\eea

The chiral expression for the operators $\bar q_i \, \overrightarrow{D}^\mu \,  L q_j$ and $\bar q_i \, \overleftarrow{D}^\mu  \, R q_j$ can be derived by using the relation: 
\bea
\bar q_i \, \overrightarrow{D}^\mu  L  \,q_j &=& \frac {1}{2} \bar q_i (\gamma^\mu \overrightarrow{\slash D} +
\overrightarrow{\slash D} \gamma^\mu ) L  \, q_j = \nn \\
& = &  - \frac {i}{2}\, m_j \bar q_i \gamma^\mu_R  \, q_j + \frac {1}{2} \partial_\nu \left( \bar q_i \gamma^\nu \gamma^\mu_L  \, q_j \right) -  \frac {1}{2} \bar q_i \overleftarrow{\slash D} \gamma^\mu_L  \, q_j = \nn \\
& = & - \frac {i}{2}\, m_j\, \bar q_i \gamma^\mu_R  \, q_j  +  \frac {1}{2} \partial^\mu \left( \bar q_i \, L  \, q_j \right) -
 \frac {1}{2} \partial_\nu \left( \bar q_i \sigma^{\mu\nu}_L  \, q_j \right) - \frac {i}{2}\, m_i\, \bar q_i \gamma^\mu_L  \, q_j \ .
\label{eq:ciuctrick}
\eea
The chiral expressions for the operators on the r.h.s of eq.\,(\ref{eq:ciuctrick}) are given in eqs.\,(\ref{eq:dim3a})-(\ref{eq:dim3f}). Note, however, that while the operator $\partial^\mu \left( \bar q_i \, L  \, q_j \right)$ is of ${\cal O}(p)$ in the chiral power counting, all other operators on the r.h.s of eq.\,(\ref{eq:ciuctrick}) are of ${\cal O}(p^3)$ and can be thus neglected. Therefore, we conclude
\bea
&& \bar q_i \, \overrightarrow{D}^\mu \,  L q_j  = - \frac{f^2}{4}B\, [\, \partial^\mu U^\dagger \,]_{ji} \, , \nn \\
&& \bar q_i \, \overleftarrow{D}^\mu \,  R q_j  = - \frac{f^2}{4}B\, [\, \partial^\mu U \,]_{ji} \ .
\label{eq:wrong}
\eea

Finally,  we also need the dimension-5 operators $\bar q_i \,  \overleftarrow{D}_\alpha  \gamma^\mu_L \,  \overrightarrow{D}^\alpha q_j$ and $g_s\, \bar q_i \,   \gamma_{\nu L} \tilde{G}^{\nu\mu} \,  q_j $. They are related by the following identity:
\bea 
\partial_\alpha \partial^\alpha  (\, \bar q_i \,  \gamma^\mu_L \, q_j ) & = & 
\partial_\alpha \left( \bar q_i \,  \overleftarrow{D}^\alpha  \gamma^\mu_L \,  q_j  +  \bar q_i \, \gamma^\mu_L \,  \overrightarrow{D}^\alpha q_j \right)  = \nn \\
&=& \bar q_i \,  \overleftarrow{D}^2  \gamma^\mu_L \,  q_j  +  \bar q_i \, \gamma^\mu_L \,  \overrightarrow{D}^2 q_j +
2 \bar q_i \,  \overleftarrow{D}_\alpha  \gamma^\mu_L \,  \overrightarrow{D}^\alpha q_j = \nn \\
&=& - 2\, g_s\, \bar q_i \,   \gamma_{\nu L} \tilde{G}^{\nu\mu} \,  q_j - (m_i^2 + m_j^2)\, \bar q_i \,   \gamma^\mu_L \,  q_j  +
 2\, \bar q_i \,  \overleftarrow{D}_\alpha  \gamma^\mu_L \,  \overrightarrow{D}^\alpha q_j \ .
\eea
Using eq.\,(\ref{eq:dim3c}) in the above expression, one obtains
\be
\bar q_i \,  \overleftarrow{D}_\alpha  \gamma^\mu_L \,  \overrightarrow{D}^\alpha q_j  -
g_s\, \bar q_i \,   \gamma_{\nu L} \tilde{G}^{\nu\mu} \,  q_j  = 
  i\, \frac{f^2}{4} \left[\, (\partial_\alpha \partial^\alpha \partial^\mu U^\dagger) \, U\,  +  
  \, (m_i^2 + m_j^2) \,  (\partial^\mu U^\dagger) \, U \right]_{ji} \, .
\label{eq:dim5}
\ee
In terms of the chiral fields one can write\,\cite{Cata:2003mn}
\be
g_s\, \bar q_i \,   \gamma_{\mu L} \tilde{G}^{\mu\nu} \,  q_j  = 
  i\, \frac{f^2}{2} \, \delta_K^2 \, \left[\, (\partial^\nu U^\dagger) \, U\, \right]_{ji} \, ,
\label{eq:semipiv}
\ee
where $\delta_K^2$ is a low-energy constant. A QCD sum rules calculation\,\cite{Cata:2003mn} provides the estimate $\delta_K^2 = 0.12 \pm 0.07 ~ {\rm GeV}^2$.
Note that the matrix  element of the operator in  eq.\,(\ref{eq:semipiv}), as all the operators of $\mathcal{O}(g_s G^A_\mu)$, is of $\mathcal{O}(\alpha^0_s )$.

We are now ready to compute the matrix elements of the 4-fermion operators in the VIA and at the lowest order in the chiral expansion. We compute always the matrix element between an initial $K^0$  and a final $\bar K^0$ state. Thus $\partial_\mu K^0 = -i p^K_\mu K^0$ and   $\partial_\mu \bar K^0 = i p^K_\mu K^0$. 
As an example, we report the detailed calculation of the matrix element of the operator $O_3$:
\bea 
&&  \langle \bar  K^0 \vert  O_3 \vert K^0\rangle = 
\langle \bar  K^0 \vert  (\bar s^a \overleftarrow{D}_\mu \, \gamma^\nu_L \, d^a  ) 
( \bar s^b \, \gamma^\nu_L \,\overrightarrow{D}^\mu  d^b )\vert K^0\rangle  = \nonumber \\ 
&= & \langle \bar  K^0 \vert  \bar s^a \overleftarrow{D}_\mu\, \gamma^\nu_L \, d^a  \vert 0\rangle 
         \langle 0 \vert \bar s^b \, \gamma^\nu_L \, \overrightarrow{D}^\mu d^b \vert K^0\rangle + 
         \langle \bar  K^0 \vert  \bar s^b \, \gamma^\nu_L \, \overrightarrow{D}^\mu d^b  \vert 0\rangle 
         \langle 0 \vert \bar s^a \overleftarrow{D}_\mu\, \gamma^\nu_L \, d^a \vert K^0\rangle + \nonumber \\ 
&+ & \langle \bar  K^0 \vert  \bar s^a \overleftarrow{D}_\mu\, \gamma^\nu_L \, \overrightarrow{D}^\mu  d^b  
         \vert 0\rangle \langle 0 \vert \bar s^b \, \gamma^\nu_L \, d^a \vert K^0\rangle +
         \langle \bar  K^0 \vert  \bar s^b \, \gamma^\nu_L \, d^a  \vert 0\rangle \langle 0 
   \vert \bar s^a\overleftarrow{D}_\mu \, \gamma^\nu_L \, \overrightarrow{D}^\mu d^b \vert K^0\rangle = \nonumber \\
&= & \langle \bar  K^0 \vert  \bar s^a \overleftarrow{D}_\mu\, \gamma^\nu_L \, d^a  \vert 0\rangle 
         \langle 0 \vert \bar s^b \, \gamma^\nu_L \, \overrightarrow{D}^\mu d^b \vert K^0\rangle + 
         \langle \bar  K^0 \vert  \bar s^b \, \gamma^\nu_L \, \overrightarrow{D}^\mu d^b  \vert 0\rangle 
         \langle 0 \vert \bar s^a \overleftarrow{D}_\mu\, \gamma^\nu_L \, d^a \vert K^0\rangle + \nonumber \\ 
&+ & \frac{1}{3} \langle \bar  K^0 \vert  \bar s^a \overleftarrow{D}_\mu\, \gamma^\nu_L \, \overrightarrow{D}^\mu  d^a  
         \vert 0\rangle \langle 0 \vert \bar s^b \, \gamma^\nu_L \, d^b \vert K^0\rangle +
         \frac{1}{3} \langle \bar  K^0 \vert  \bar s^a \, \gamma^\nu_L \, d^a  \vert 0\rangle \langle 0 
         \vert \bar s^b \overleftarrow{D}_\mu \, \gamma^\nu_L \, \overrightarrow{D}^\mu d^b \vert K^0\rangle = \nonumber \\
&=& f_K^2 \,  \left[ \frac{1}{2}\, M_K^4  - \frac{1}{8}\, B^2 \left(m_s-m_d\right)^2 \right]  - 
         \frac{1}{3}\, f_K^2\, M_K^2 \left[ M_K^2 - \left( m_s^2+m_d^2 \right) - 2\, \delta_K^2 \right]  = \nonumber \\
&=&  \frac{1}{24}\, f_K^2 \,  \left[ 4\, M_K^4  - 3\, B^2 \left(m_s-m_d\right)^2  + 
         8\, M_K^2 \left( m_s^2+m_d^2 \right) + 16\, M_K^2 \, \delta_K^2 \right] \, ,
\eea
where, in evaluating the matrix element between kaon states we have replaced the low energy constant $f$ with the kaon decay constant $f_K$. By neglecting the mass of the down quark, $m_d$, and thus also replacing $B = M_K^2/m_s$, one finally obtains
\be
\langle \bar  K^0 \vert  O_3 \vert K^0\rangle = \frac{1}{24}\, f_K^2 \,  M_K^2\, \left[ M_K^2  + 8\, m_s^2  + 16 \, \delta_K^2 \right]\,.
\ee

The VIA for the matrix elements of the other relevant 4-fermion operators can be derived in a similar way. In the limit $m_d \to 0$, we obtain
\bea
\langle \bar  K^0 \vert  O_1 \vert K^0\rangle & = & \frac{5}{3}\, f_K^2 \,  M_K^4 \, , \nn \\
\langle \bar  K^0 \vert  O_2 \vert K^0\rangle & = & \frac{11}{12}\, f_K^2 \,  M_K^4 \, , \nn \\
\langle \bar  K^0 \vert  O_3 \vert K^0\rangle & = & \frac{1}{24}\, f_K^2 \,  M_K^2\, \left[ 16 \, \delta_K^2 + M_K^2  + 8\, m_s^2  \right] \, , \nn \\
\langle \bar  K^0 \vert  O_4 \vert K^0\rangle & = & 2\, f_K^2 \,  M_K^2 \delta_K^2 \, , \\
\langle \bar  K^0 \vert  O_5 \vert K^0\rangle & = & - f_K^2 \,  M_K^4 \, , \nn \\
\langle \bar  K^0 \vert  O_6 \vert K^0\rangle & = & \frac{8}{3}\, f_K^2 \,  M_K^2 \, m_s^2 \, , \nn \\
\langle \bar  K^0 \vert  O_7 \vert K^0\rangle & = & - \frac{5}{3}\, f_K^2 \,  M_K^4 \, , \nn \\
\langle \bar  K^0 \vert  O_8 \vert K^0\rangle & = & 0 \, . \nn
\eea
Note that in the chiral power counting the leading contribution is given by the operators $O_3$ and $O_4$ whose matrix elements are of ${\cal O}(p^2)$ (for $O_3$, however, the ${\cal O}(p^2)$ contribution is suppressed by $1/N_c$). The matrix element of  $O_6$ is of ${\cal O}(p^6)$ while all other matrix elements are of ${\cal O}(p^4)$.
In ref.\,\cite{Cata:2004ti} the chiral expansion was performed at the leading order and the large-$N_c$ limit was  assumed. As a consequence, only the $O_4$ contribution was included.
We observe that the contribution of the other seven operators, which we include in the present work for the first time, can a priori introduce an $\mathcal{O}(1)$ correction to the $O_4$ contribution, as $m_K$ is numerically comparable  to  $\delta_K$ ($\delta_K^2/M_K^2\simeq 0.5$).

\section{A possible approach to compute non-local contributions at \texorpdfstring{$\mathcal{O}(\alpha_s p^2/m_c^2)$}{order alphas p2 over mc2}}
\label{sec:possible}
Some   remark  may be useful at this point.  Concerning the non-local contributions that appear at $\mathcal{O}(\alpha_sp^2/m_c^2)$, let us take, for example,    the diagram on the left upper panel  of  fig.\,\ref{fig:penaprox}
at the lowest perturbative order in the Standard Model, by considering  only the CP violating  contribution to   $K^0 - \bar K^0$ mixing. By   using unitarity ($\lambda_u+\lambda_c+\lambda_t=0$),   neglecting  $\lambda_t  \times  \mathcal{O}(p^2/m_c^2)$ corrections and  without the resummation of large QCD logarithms,   we find  that  there  is only  one contribution to the amplitude which goes as 
\be  \mathcal{A}(\Delta S=1 \otimes \Delta S=1)  \sim \frac{G_F^2}{16\, \pi^2}\, \mathrm{Im}(\lambda_c^2)  \,  \frac{\alpha_s}{4\pi}\,\frac{4}{3} \,\log\left(\frac{m_t^2}{m_c^2}\right) \, \left(\Pi[p^2, m_u^2]-\Pi[p^2, m_c^2]\right)\, , \label{eq:effamp} \ee
where $\alpha_s$ is the strong coupling constant  and  the factor  $\log\left(m_t^2/m_c^2\right)$  corresponds to the Wilson coefficient of the    operator 
\be  Q= \bar s t^A \gamma^\mu_L d  \, \sum_q \bar q t^A \gamma_\mu q \, , \label{eq:QGIM} \ee
 generated  by the  heavy charm-top quark GIM  penguin.   $\Pi[p^2, m_{u}^2]-\Pi[p^2, m_{c}^2]$  denotes the hadronic contribution due the contraction of the operator  $Q$ with the  operator
\be  Q_2=Q^u_2-Q^c_2=\bar s  \gamma^\mu_L u  \,\bar u  \gamma^L_\mu d-\bar s  \gamma^\mu_L c  \,\bar c  \gamma^L_\mu d\, , \ee
  which appears at tree level in the effective weak Hamiltonian.    $\Pi[p^2, m_{u}^2]$ corresponds to the diagram on the upper right panel of fig.\,\ref{fig:penaprox} with (light)  up quarks  in  the  loop and must be  computed non-perturbatively; 
    $\Pi[p^2, m_{c}^2]$ denotes the  same diagram with   charm-quarks  in the loop and,  whithin the  present approximation, can be computed in perturbation  theory with a suitable matching to the theory with a propagating charm. 
  GIM    is essential to make the final result finite at the lowest non-trivial order in  the strong coupling constant. The combination  $\left(\Pi[p^2, m_u^2]-\Pi[p^2, m_c^2]\right)$ is of $\mathcal{O}(p^2)$ and, as expected, it is a correction of $\mathcal{O}(\alpha_s p^2/m_c^2)$ to $\varepsilon_K$  since the leading term in Im$(\lambda_c^2)$, coming from the box diagram of fig. 1, goes as $G_F^2/(16 \pi^2) m_c^2$.

Thus, by expanding at the leading order in $1/m_t^2$ and $1/m_c^2$ and by  using the equations of motion, we may transform the charm-top quark GIM  penguin of the Feynman diagram  on the left upper panel of   fig.\,\ref{fig:penaprox}  in the insertion  of the  operator  $Q$  
multiplied by its Wilson coefficient, thus obtaining the Feynman diagram on the r.h.s. of  upper panel of  fig.\,\ref{fig:penaprox}, which is a standard {\it connected} diagram  much easier to evaluate in lattice QCD~\,\cite{Christ:2014qwa,Bai:2018mdv}. 
Following the standard approach, even if the  momentum flowing in the charm-top GIM  penguin is of  $\mathcal{O}(p^2) \sim \mathcal{O}(\Lambda_{QCD}^2)$,   in  eq.\,(\ref{eq:effamp}) the coefficient of    $Q$  can  be computed in perturbation theory  since  both the quarks in the GIM loop are heavy (they are either the top  or the charm).
  Eventually,  the most convenient  choice  is to  take  $\mu =m_c$  as renormalisation scale.   In  the calculation of the double penguins, in  order take into account higher order QCD  corrections,    we  can imagine to substitute to the Standard Model the double insertion  of the low energy $\Delta S=1$  Hamiltonian and compute  on the lattice,  with this effective theory,  the charm-top  and up-charm GIM penguins. This extreme  accuracy, although feasible, is probably not necessary since we are  evaluating one of the  corrections to the  leading term and the uncertainties entailed in other terms, specifically  in the calculation of the   matrix elements of the local dimension-8  operators,  are probably of  the size, if not larger,  than the  neglected higher-order QCD effects.    The calculation of the other double penguin diagram, shown on the  left  lower panel of   fig.\,\ref{fig:penaprox}  could  proceed in the same way by replacing it  with the    connected diagram on the right lower panel of the same figure.


\boldmath
\section{Result for the power corrections to \texorpdfstring{$\varepsilon_K$}{epsilonK}}
\unboldmath
\label{sec:prospects}
By combining the results for the Wilson coefficients of the effective Hamiltonian given in section 3 and those for the matrix elements of the four fermion operators evaluated in section 4 in the VIA, we obtain
\bea
\langle \bar  K^0 \vert  H^{ \mathrm{eff}}_{1/m_c^2} \vert K^0\rangle &=& 
 \frac{G_F^2}{16\, \pi^2} \, \mathrm{Im}\left(\lambda_c^2\right) \,\sum_i C_i \, \langle \bar  K^0 \vert  O_i \vert K^0\rangle = \nn\\
 &=& \frac{G_F^2}{16\, \pi^2} \, \mathrm{Im}\left(\lambda_c^2\right) \, \left[ - \dfrac{1}{108} \, f_K^2 \,  M_K^2\, \left( 272\, \delta_K^2 + 15 \, M_K^2 \right)\right] \ ,
\label{eq:main}
\eea
which represents the main result of this paper. In eq.\,(\ref{eq:main}) we have neglected small terms of ${\cal O}(p^6)$, proportional to $M_K^2\, m_s^2$, that contribute for $\sim 1.5\%$ of the total correction. The above result can be compared with the leading\footnote{The effective Hamiltonian in eq.\,(\ref{eq:leading}) is at the leading order in the OPE,  while it includes  the NLO QCD and EW corrections, which are enclosed in the $\eta$'s\,\cite{Buras:1990fn,Herrlich:1993yv,Herrlich:1996vf,Brod:2021qvc}.} contribution to the effective Hamiltonian that can be written in the form
\bea
\langle \bar  K^0 \vert  H^{ \mathrm{eff}}_{LT} \vert K^0\rangle &=& \frac{G_F^2}{16\, \pi^2} \, M_W^2 \, \left[ 
 \mathrm{Im}\left(\lambda_c^2\right) \, \left( \eta_1 S_0(x_c) - \eta_3 S_0(x_c,x_t) \right) + \right. \nn \\
&& \left. +  \mathrm{Im}\left(\lambda_t^2\right) \, \left( \eta_2 S_0(x_t) - \eta_3 S_0(x_c,x_t) \right) \right] \, \dfrac{8}{3} \, f_K^2 \,  M_K^2 \, \hat B_K \ .
\label{eq:leading}
\eea
Therefore, with respect to the term proportional to $ \mathrm{Im}\left(\lambda_c^2\right)$ in the LT effective Hamiltonian, $\langle \bar  K^0 \vert  H^{ \mathrm{eff}}_{LT} \vert K^0\rangle_{ \mathrm{Im}\left(\lambda_c^2\right)}$,   eq.\,(\ref{eq:main}) represents a relative correction given by 
\bea
\frac{\langle \bar  K^0 \vert  H^{ \mathrm{eff}}_{1/m_c^2} \vert K^0\rangle}{\langle \bar  K^0 \vert  H^{ \mathrm{eff}}_{LT} \vert K^0\rangle_{ \mathrm{Im}\left(\lambda_c^2\right)}}&=&
\left[\left( \eta_1 \dfrac{S_0(x_c)}{x_c} - \eta_3 \dfrac{S_0(x_c,x_t)}{x_c} \right) \hat B_K \right]^{-1} \,
\left(-\dfrac{17}{18}\, \dfrac{\delta_K^2}{m_c^2} - \dfrac{5}{96} \, \dfrac{M_K^2}{m_c^2} \right) \, \simeq \nn \\
&\simeq & -0.51 \, \left(-\dfrac{17}{18}\, \dfrac{\delta_K^2}{m_c^2} - \dfrac{5}{96} \, \dfrac{M_K^2}{m_c^2} \right) \simeq
0.04 \,.
\eea
Up to higher order terms, we may then write 
\bea \mathrm{Im} M_{12}=  \langle \bar  K^0 \vert  H^{ \mathrm{eff}}_{LT} \vert K^0\rangle \, \left(1 +\delta_{BGI}+\delta_{m_c}\right)\, , \eea
where $\delta_{BGI}=  0.02$    was computed in ref.\,\cite{Buras:2010pza} and
the relative correction with respect to the full LT Hamiltonian is given by 
\bea
\delta_{m_c} &\equiv& \frac{\langle \bar  K^0 \vert  H^{ \mathrm{eff}}_{1/m_c^2} \vert K^0\rangle}{\langle \bar  K^0 \vert  H^{ \mathrm{eff}}_{LT} \vert K^0\rangle}\,=\nn\\
&=&\left[\left( \eta_1 \dfrac{S_0(x_c)}{x_c} - \eta_3 \dfrac{S_0(x_c,x_t)}{x_c} \right) - \dfrac{Re(\lambda_t)}{Re(\lambda_c)}
\left( \eta_2 \dfrac{S_0(x_t)}{x_c} - \eta_3 \dfrac{S_0(x_c,x_t)}{x_c} \right) \right]^{-1} \cdot \label{eq:numerical}
 \\
&\cdot& \hat B_K^{-1} \, \left(-\dfrac{17}{18}\, \dfrac{\delta_K^2}{m_c^2} - \dfrac{5}{96} \, \dfrac{M_K^2}{m_c^2} \right) \, \simeq 
-0.13 \, \left(-\dfrac{17}{18}\, \dfrac{\delta_K^2}{m_c^2} - \dfrac{5}{96} \, \dfrac{M_K^2}{m_c^2} \right) \simeq 0.010 \pm 0.003\,, \nonumber
\eea that is the $\delta_{m_c}$ correction increases the theoretical
prediction for $\varepsilon_K$ by 1\%. As is evident from
eq.\,(\ref{eq:numerical}), the $\mathcal{O}(m_K^2/m_c^2)$ correction,
included here for the first time, though a priori comparable to the
$\mathcal{O}(\delta_K^2/m_c^2)$ term, turns out to be numerically
smaller. At the origin of this numerical result there is the larger
numerical value of the Wilson coefficient $C_4$ w.r.t the coefficients
of the other seven operators, and some cancellation occurring among
the terms proportional to $f_K^2 \, m_K^4$. The dominant uncertainty
on our result for $\delta_{m_c}$ is represented by the VIA. We evaluated it
assuming for the matrix elements of the dimension-8 operators either a
uniform distribution centered on the VIA with a $50\%$ half width or a Gaussian distribution with $\sigma = 30\%$; in both cases we obtain a $30\%$ error on
$\delta_{m_c}$, which keeps a negligible impact on the $\varepsilon_K$ total
uncertainty.

We conclude by providing an updated theoretical
prediction for $\varepsilon_K$, which includes our result for
$\delta_{m_c}$ of eq.\,(\ref{eq:numerical}), the short- and
long-distance contributions recently calculated in
refs.~\cite{Buras:2008nn}, \cite{Buras:2010pza}, \cite{Brod:2021qvc} and
uses for the bag-parameter the value $\hat B_K=0.756 \pm 0.016$,
obtained by averaging the $N_f = 2+1$ and $N_f=2+1+1$ FLAG
averages\,\cite{Aoki:2019cca}. Our prediction is obtained
  using the UTfit code, in the context of an updated Bayesian
  Unitarity Triangle Analysis \cite{UTfitSummer21}, which yields
  (omitting of course the constraint from $\varepsilon_K$)
  Im$(\lambda_t)=(13.44\pm 0.57)\cdot 10^{-5}$,
  Re$(\lambda_t)=(-3.11\pm 0.11)\cdot 10^{-4}$ and
  Re$(\lambda_c)=(-2191.0\pm 7.5)\cdot 10^{-4}$. All the relevant inputs can be found in \cite{UTfitSummer21}. Our theoretical
prediction for $\varepsilon_K$ reads
\begin{equation}  \varepsilon_K =  (1.99 \pm  0.14) \times10^{-3}\, .\end{equation}

\vspace*{1.cm}

\section*{Acknowledgments}
We gratefully acknowledge M. Bona, C.T. Sachrajda  and   M. Gorbahn for useful discussions on higher perturbative orders.
We thank MIUR  (Italy)  for  partial  support  under  the  contracts  PRIN 2015 protocollo 2015P5SBHT and PRIN 20172LNEEZ. 

\section*{Appendix A: Useful relations coming from the equations of motion}
In the calculation of the contribution of the gluon on an external leg, both in the full and in the effective theory, we can eliminate some structures containing the completely antisymmetric tensor, by using some relations originated from the equations of motion.
These relations are obtained from the Chisholm identity
\be
\gamma^{\alpha} \gamma^{\beta} \gamma^{\gamma}_L=g^{\alpha \beta} \gamma^{\gamma}_L-g^{\alpha \gamma} \gamma^{\beta}_L+g^{\beta \gamma} \gamma^{\alpha}_L-i\,\epsilon^{\alpha \beta \gamma \delta} \gamma^{\delta}_L\,,
\ee
by saturating one Lorentz index with a momentum which induces the Dirac equation of motion.
These relations show that four of the structures involving  the completely antisymmetric tensor can be reduced in terms of other structures.
They can be written as
\bea
\epsilon^{\alpha \nu \beta \rho} p_1^\beta \bar s(p_2)\gamma^\alpha_L d(p_1) &=& i p_1^\nu \bar s(p_2)\gamma^\rho_L d(p_1) -i p_1^\rho \bar s(p_2)\gamma^\nu_L d(p_1)+i m_d \bar s(p_2)\sigma^{\nu\rho}_R d(p_1)\,,\\
\epsilon^{\alpha \nu \beta \rho} p_2^\beta \bar s(p_2)\gamma^\alpha_L d(p_1) &=& -i p_2^\nu \bar s(p_2)\gamma^\rho_L d(p_1) +i p_2^\rho \bar s(p_2)\gamma^\nu_L d(p_1)-i m_s \bar s(p_2)\sigma^{\nu\rho}_L d(p_1)\,,\\
\epsilon^{\alpha \nu \gamma \rho} p_3^\gamma \bar s(p_3)\gamma^\nu_L d(p_4) &=& i p_3^\alpha \bar s(p_3)\gamma^\rho_L d(p_4) -i p_3^\rho \bar s(p_3)\gamma^\alpha_L d(p_4)-i m_s \bar s(p_3)\sigma^{\alpha\rho}_L d(p_4)\,,\\
\epsilon^{\beta \gamma \mu \nu} p_2^\gamma \bar s(p_3)\gamma^\beta_L d(p_4) &=& -i (p_1+p_2-2p_3-q)^\mu \bar s(p_3)\gamma^\nu_L d(p_4) +i (p_1+p_2-2p_3-q)^\nu \bar s(p_3)\gamma^\mu_L d(p_4)\nn\\
&&-i m_s \bar s(p_3)\sigma^{\mu\nu}_L d(p_4)+i m_d \bar s(p_3)\sigma^{\mu\nu}_R d(p_4)\nn\\
&&+\epsilon^{\beta \gamma \mu \nu} q^\gamma \bar s(p_3)\gamma^\beta_L d(p_4)-\epsilon^{\beta \gamma \mu \nu} p_1^\gamma \bar s(p_3)\gamma^\beta_L d(p_4)\,.
\eea

\section*{Appendix B: Useful Fierz relations}
The matrix elements of the operators entering the effective Hamiltonian $H^{ \mathrm{eff}}_{1/m_c^2}$ of eq.\,(\ref{eq:Heff1mc2}) are estimated by using the VIA, in which Fierz transformations play an important role. The Fierz transformations can be obtained starting from the following general formula
\begin{eqnarray}
\Gamma_1 \,\otimes\, \Gamma_2 & \xrightarrow{Fierz} &  
\frac{1}{8}\, \Gamma_1  R \Gamma_2 \,\otimes\, R \ + \
\frac{1}{8}\, \Gamma_1  L \Gamma_2 \,\otimes\, L \ + \nonumber \\ & + &
\frac{1}{8}\, \Gamma_1  \gamma_R^\alpha \Gamma_2 \,\otimes\, \gamma_L^\alpha + 
\frac{1}{8}\, \Gamma_1  \gamma_L^\alpha \Gamma_2 \,\otimes\, \gamma_R^\alpha - \nonumber \\ & - &
\frac{1}{32}\, \Gamma_1  \sigma_L^{\alpha\beta} \Gamma_2 \,\otimes\, \sigma_L^{\alpha\beta}  \ - \
\frac{1}{32}\, \Gamma_1  \sigma_R^{\alpha\beta} \Gamma_2 \,\otimes\, \sigma_R^{\alpha\beta} \ ,
\label{eq:fierz_ch}
\end{eqnarray}
where we used the definition $\sigma^{\alpha\beta} = \left[\gamma^\alpha , \gamma^\beta \right]/2$. 
From eq.\,(\ref{eq:fierz_ch}) one derives, in particular,
\begin{equation}
\gamma^\mu_L \,\otimes\, \gamma^\nu_L \xrightarrow{Fierz}
\frac{1}{2}\, \gamma^\mu_L \,\otimes\, \gamma^\nu_L \ + \
\frac{1}{2}\, \gamma^\nu_L \,\otimes\, \gamma^\mu_L \  - \
\frac{1}{2}\, g^{\mu\nu} \, \gamma^\rho_L \,\otimes\, \gamma^\rho_L - 
\frac{i}{2}\, \varepsilon^{\mu\nu\rho\sigma} \, \gamma^\rho_L \,\otimes\, \gamma^\sigma_L
\label{eq:fs1}
\end{equation}
\begin{equation}
L \,\otimes\, \gamma^\mu_L \xrightarrow{Fierz}
\frac{1}{2}\, L \,\otimes\, \gamma^\mu_L \ - \
\frac{1}{2}\, \sigma^{\mu\nu}_L \,\otimes\, \gamma^\nu_L
\end{equation}
\begin{equation}
R \,\otimes\, \gamma^\mu_L \xrightarrow{Fierz}
\frac{1}{2}\, \gamma^\mu_L \,\otimes\, R \ + \
\frac{1}{2}\, \gamma^\nu_L \,\otimes\, \sigma^{\mu\nu}_R
\end{equation}

It is also useful to report the general formula for the Fierz of color indices. It reads:
\be
C_1 \,\otimes\, C_2 \xrightarrow{Fierz} 
\dfrac{1}{N}\, C_1\, C_2 \,\otimes\, I + 2\, C_1\, t^A \, C_2 \,\otimes\, t^A \ ,
\label{eq:fierz_color}
\ee
where $C_1$ and $C_2$ are generic color matrices. From eq.\,(\ref{eq:fierz_color}) one derives, in particular,
\be
t^A \,\otimes\, I - I \, \otimes \, t^A \xrightarrow{Fierz}  - 2\, i\, f_{ABC} \, t^B \,\otimes\, t^C \ .
\label{eq:fc1}
\ee
\section*{Appendix  C:   Matching with the external state made up of four quarks and a gluon}
In this appendix we collect the amplitudes that enter the matching with the external state made up of four quarks and a gluon.
As discussed in the main text for the case of the four quark external state, one can choose equal color indices for the quark and anti-quark in the initial state and similarly for the final state, which corresponds to selecting the box diagram in fig.\,\ref{fig:boxes} (a). 
We do the same also when the gluon is added in the external state. With the additional gluon, moreover, one can impose the matching condition by considering only contributions with the color generator $t^A$ appearing in the initial state bilinear or in the final state bilinear.
We perform the first choice, that is we consider the amplitudes having the form
\be
\bar s^a(p_2) \Gamma^i \, t^A_{a\,b} \,d^b(p_1) \bar s^c(p_3) \Gamma^j d^c(p_4)\,.
\ee

 By omitting for brevity the external spinors (and the gluon polarization vector), the matrix elements for the various operators in eq.\,(\ref{eq:pivovarov}), read
\bea
 \langle O_{1} \rangle_{4q1g}  &=& i\, \gamma^{\alpha}_L\,t^A \otimes \gamma^{\nu}_L \,\epsilon^{\alpha\, \nu\, \sigma\, \rho }\, q^{\sigma}\left( \frac{ \left(m_d^2+m_s^2-4 p_2.q+2 p_1.p_2-2 p_1.p_3\right)}{p_2.q}\right.\nn\\
  && \left. + \frac{\left(m_d^2+m_s^2-2 p_2.q+2 p_3.q+2 p_1.p_2-2 p_1.p_3\right)}{p_1.q}-2 \,\frac{p_1.q}{ p_2.q} \right)\nn\\
   && + \gamma^{\mu}_L\,t^A \otimes \gamma^{\mu}_L \left(-\frac{2\, p_1^{\rho } \left(m_d^2+m_s^2-2 p_1.q-2 p_2.q+2 p_3.q+2 p_1.p_2-2 p_1.p_3\right)}{p_1.q}\right.\nn\\
   && \left.+\frac{2\, p_2^{\rho } \left(m_d^2+m_s^2-2 p_1.q-2 p_2.q+2 p_1.p_2-2 p_1.p_3\right)}{p_2.q}+4 p_3^{\rho }\right)\nn\\
   && +\gamma^{\rho}_L\,t^A \otimes \cancel{q}_L \left(-\frac{m_d^2+m_s^2+2 \,p_2.q+2\, p_1.p_2-2\, p_1.p_3}{p_2.q}\right.\nn\\
   && \left. +\frac{m_d^2 +m_s^2-2\, p_2.q+2\, p_3.q+2\, p_1.p_2-2\, p_1.p_3}{p_1.q}  +\frac{2\, p_1.q}{p_2.q}\right)\nn\\
   && +\cancel{q}_L\,t^A \otimes \gamma^{\rho}_L \left(\frac{m_d^2+m_s^2+2\, p_2.q+2\, p_1.p_2-2\, p_1.p_3}{p_2.q}\right.\nn\\
   &&\left. -\frac{m_d^2+m_s^2-2\, p_2.q+2\, p_3.q+2\, p_1.p_2-2\, p_1.p_3}{p_1.q}-\frac{2\, p_1.q}{p_2.q}\right)\,,\nn\\\\ \langle O_{2} \rangle_{4q1g}  &=& -\frac{i\, \gamma^{\alpha}_L \, t^A \otimes R\, m_d \left(p_1.q+p_2.q\right) \epsilon^{\alpha\, \nu\, \sigma\, \rho }\, p_3^{\nu}\, q^{\sigma}}{p_1.q\,p_2.q} \nn\\
&&+\frac{i \,\gamma^{\alpha}_L\,t^A \otimes L\, m_s \left(p_1.q+p_2.q\right)
   \epsilon^{\alpha\, \nu\, \sigma\, \rho }\, p_3^{\nu}\, q^{\sigma}}{p_1.q\, p_2.q}\nn\\
   && -\frac{2\, i\, \gamma^{\alpha}_L \, t^A \otimes \cancel{p_1}_L
   \left(p_1.q+p_2.q\right) \epsilon^{\alpha\, \nu\, \sigma\, \rho }\, p_3^{\nu}\, q^{\sigma}}{p_1.q\,
   p_2.q} +\frac{2\, i\, \gamma^{\alpha}_L\,t^A \otimes \cancel{q}_L\, \epsilon^{\alpha\, \nu\, \sigma\, \rho }\, p_3^{\nu}\, q^{\sigma}}{p_1.q}\nn\\
   && +\cancel{p_3}_L \, t^A \otimes R \left(\frac{2\, m_d\, p_1^{\rho }}{p_1.q}-\frac{2\, m_d\, p_2^{\rho }}{p_2.q}\right)+\cancel{q}_L \, t^A \otimes R\, m_d \left(\frac{1}{p_1.q}-\frac{1}{p_2.q}\right) p_3^{\rho }\nn\\
   &&+\gamma^{\rho}_L \, t^A \otimes R\, m_d \left(\frac{1}{p_2.q}-\frac{1}{p_1.q}\right) p_3.q+R \, t^A \otimes \cancel{p_1}_L
   \left(\frac{2\, m_d\, p_2^{\rho }}{p_2.q}-\frac{2\, m_d\, p_1^{\rho }}{p_1.q}\right)\nn\\
   &&+q^{\nu}\,\sigma^{\nu\,\rho}_R \, t^A \otimes \cancel{p_1}_L\, m_d
   \left(\frac{1}{p_2.q}+\frac{1}{p_1.q}\right)\nn +\frac{2\, R \, t^A \otimes \cancel{q}_L\, m_d\, p_1^{\rho
   }}{p_1.q}-\frac{q^{\nu}\, \sigma^{\nu\,\rho}_R\,t^A \otimes \cancel{q}_L\, m_d}{p_1.q}\nn\\
   &&-2\, R\,t^A \otimes \gamma^{\rho}_L\, m_d+L \, t^A \otimes \cancel{p_1}_L \left(\frac{2 \,m_s\,
   p_1^{\rho }}{p_1.q}-\frac{2\, m_s\, p_2^{\rho }}{p_2.q}\right) +q^{\nu}\,\sigma^{\nu\,\rho}_L \, t^A \otimes \cancel{p_1}_L
   \left(-\frac{m_s}{p_1.q}-\frac{m_s}{p_2.q}\right)\nn\\
   &&-\frac{2\, L \, t^A \otimes \cancel{q}_L\, m_s\, p_1^{\rho
   }}{p_1.q}+\frac{q^{\nu}\,\sigma^{\nu\,\rho}_L\,t^A \otimes \cancel{q}_L \,m_s}{p_1.q}+\cancel{p_3}_L\,t^A \otimes L \left(\frac{2\, m_s\, p_2^{\rho
   }}{p_2.q}-\frac{2\, m_s\, p_1^{\rho }}{p_1.q}\right)\nn\\
   &&+\cancel{q}_L\, t^A \otimes L\, m_s
   \left(\frac{1}{p_2.q}-\frac{1}{p_1.q}\right) p_3^{\rho } +\gamma^{\rho}_L \, t^A \otimes L \,m_s
   \left(\frac{1}{p_1.q}-\frac{1}{p_2.q}\right) p_3.q+2\, L\,t^A \otimes \gamma^{\rho}_L\, m_s \nn\\
   &&+\cancel{p_3}_L \, t^A \otimes \cancel{p_1}_L \left(\frac{4
   p_1^{\rho }}{p_1.q}-\frac{4\, p_2^{\rho }}{p_2.q}\right) +2\, \cancel{q}_L \, t^A \otimes \cancel{p_1}_L
   \left(\frac{1}{p_1.q}-\frac{1}{p_2.q}\right) p_3^{\rho }\nn\\
   &&+2 \,\gamma^{\rho}_L\,t^A \otimes \cancel{p_1}_L
   \left(\frac{1}{p_2.q}-\frac{1}{p_1.q}\right) p_3.q -\frac{4 \,\cancel{p_3}_L\,t^A \otimes \cancel{q}_L\, p_1^{\rho
   }}{p_1.q}-\frac{2\, \cancel{q}_L\,t^A \otimes \cancel{q}_L\, p_3^{\rho }}{p_1.q}\nn\\
   &&+\gamma^{\rho}_L\,t^A \otimes \cancel{q}_L \left(\frac{2
   p_3.q}{p_1.q}-2\right)+4\, \cancel{p_3}_L\,t^A \otimes \gamma^{\rho}_L+2\, \cancel{q}_L\,t^A \otimes \gamma^{\rho}_L\,,\nn\\\\
 \langle O_{3} \rangle_{4q1g}  &=&\frac{i\, \gamma^{\alpha}_L\,t^A \otimes \gamma^{\nu}_L \left(p_1.p_3 \left(p_1.q+p_2.q\right)-p_2.q\, p_3.q\right)
   \epsilon^{\alpha\, \nu\, \sigma\, \rho }\, q^{\sigma}}{p_1.q\, p_2.q}\nn\\
 && +\gamma^{\mu}_L\,t^A \otimes \gamma^{\mu}_L \left(\frac{2\, p_3.q\, p_1^{\rho
   }}{p_1.q}+p_1.p_3 \left(\frac{2\, p_2^{\rho }}{p_2.q}-\frac{2\, p_1^{\rho }}{p_1.q}\right)-2\,
   p_3^{\rho }\right) \nn\\
   && +\gamma^{\rho}_L\,t^A \otimes \cancel{q}_L \left(p_1.p_3 \left(\frac{1}{p_1.q}-\frac{1}{p_2.q}\right)-\frac{p_3.q}{p_1.q}+1\right)\nn\\
   &&+\cancel{q}_L\,t^A \otimes \gamma^{\rho}_L \left(p_1.p_3 \left(\frac{1}{p_2.q}-\frac{1}{p_1.q}\right)+\frac{p_3.q}{p_1.q}-1\right)\, ,\nn\\\\
 \langle O_{4} \rangle_{4q1g}  &=& i \,\gamma^{\alpha}_L\,t^A \otimes \gamma^{\nu}_L\, \epsilon^{\alpha\, \nu\, \sigma\, \rho }\, q^{\sigma}\,,\nn\\\\
 \langle O_{5} \rangle_{4q1g}  &=& \frac{i\, \gamma^{\alpha}_L \, t^A \otimes R\, m_d \left(p_1.q+p_2.q\right) \epsilon^{\alpha\, \nu\, \sigma\, \rho }\, p_3^{\nu}\, q^{\sigma}}{2\, p_1.q\, p_2.q}\nn\\
&&+\frac{i\, \gamma^{\alpha}_L\,t^A \otimes L\, m_s \left(p_1.q+p_2.q\right)
   \epsilon^{\alpha\, \nu\, \sigma\, \rho }\, p_3^{\nu}\, q^{\sigma}}{2\, p_1.q\, p_2.q}\nn\\
   && +q^{\nu}\, \sigma^{\nu\,\rho}_R \, t^A \otimes L\, m_d\, m_s
   \left(-\frac{1}{p_2.q}-\frac{1}{p_1.q}\right)+2\, R\, t^A \otimes L\, m_d\, m_s \left(\frac{p_1^{\rho
   }}{p_1.q}-\frac{p_2^{\rho }}{p_2.q}\right)\nn\\
   && +R \, t^A \otimes R\, m_d^2 \left(\frac{p_2^{\rho
   }}{p_2.q}-\frac{p_1^{\rho }}{p_1.q}\right)+\cancel{p_3}_L \, t^A \otimes R \left(\frac{m_d\, p_2^{\rho
   }}{p_2.q}-\frac{m_d\, p_1^{\rho }}{p_1.q}\right) \nn\\
   && +\frac{1}{2} \cancel{q}_L \, t^A \otimes R\, m_d
   \left(\frac{1}{p_2.q}-\frac{1}{p_1.q}\right) p_3^{\rho }+\frac{1}{2} \gamma^{\rho}_L \, t^A \otimes R\, m_d
   \left(\frac{1}{p_1.q}-\frac{1}{p_2.q}\right) p_3.q\nn\\
   &&+\frac{q^{\nu}\,\sigma^{\nu\,\rho}_R \, t^A \otimes R\, m_d^2
   \left(p_1.q+p_2.q\right)}{2\, p_1.q\, p_2.q} +R \, t^A \otimes \cancel{p_1}_L \left(\frac{m_d\, p_2^{\rho
   }}{p_2.q}-\frac{m_d\, p_1^{\rho }}{p_1.q}\right)\nn\\
   &&+\frac{1}{2} q^{\nu}\,\sigma^{\nu\,\rho}_R \, t^A \otimes \cancel{p_1}_L\, m_d
   \left(\frac{1}{p_2.q}+\frac{1}{p_1.q}\right)+\frac{R \, t^A \otimes \cancel{q}_L\, m_d\, p_1^{\rho
   }}{p_1.q}-\frac{q^{\nu}\, \sigma^{\nu\,\rho}_R\,t^A \otimes \cancel{q}_L\, m_d}{2 \,p_1.q}\nn\\
   &&-R\,t^A \otimes \gamma^{\rho}_L\, m_d+\frac{q^{\nu}\, \sigma^{\nu\,\rho}_L \, t^A \otimes L\, m_s^2
   \left(p_1.q+p_2.q\right)}{2\, p_1.q\, p_2.q}+L \, t^A \otimes \cancel{p_1}_L \left(\frac{m_s\, p_2^{\rho
   }}{p_2.q}-\frac{m_s\, p_1^{\rho }}{p_1.q}\right)\nn\\
   &&+\frac{1}{2} q^{\nu}\,\sigma^{\nu\,\rho}_L \, t^A \otimes \cancel{p_1}_L\, m_s
   \left(\frac{1}{p_2.q}+\frac{1}{p_1.q}\right)+\frac{L \, t^A \otimes \cancel{q}_L\, m_s\, p_1^{\rho
   }}{p_1.q}\nn\\
   &&+L\,t^A \otimes L\, m_s^2 \left(\frac{p_2^{\rho }}{p_2.q}-\frac{p_1^{\rho
   }}{p_1.q}\right) -\frac{q^{\nu}\,\sigma^{\nu\,\rho}_L\,t^A \otimes \cancel{q}_L \,m_s}{2\, p_1.q}+\cancel{p_3}_L\,t^A \otimes L \left(\frac{m_s\, p_2^{\rho
   }}{p_2.q}-\frac{m_s\, p_1^{\rho }}{p_1.q}\right)\nn\\
   && +\frac{1}{2} \cancel{q}_L\, t^A \otimes L\, m_s
   \left(\frac{1}{p_2.q}-\frac{1}{p_1.q}\right) p_3^{\rho }+\frac{1}{2} \gamma^{\rho}_L \, t^A \otimes L\, m_s
   \left(\frac{1}{p_1.q}-\frac{1}{p_2.q}\right) p_3.q-L\,t^A \otimes \gamma^{\rho}_L\, m_s\,,\nn\\\\
 \langle O_{6} \rangle_{4q1g}  &=& \frac{i \,\gamma^{\alpha}_L\,t^A \otimes \gamma^{\nu}_L \left(p_1.q+p_2.q\right) \left(m_d^2+m_s^2\right) \epsilon^{\alpha\, \nu\, \sigma\, \rho }\, q^{\sigma}}{p_1.q\, p_2.q}\nn\\
  && +\frac{2 \,\gamma^{\mu}_L\,t^A \otimes \gamma^{\mu}_L \left(m_d^2+m_s^2\right) \left(p_1.q \, p_2^{\rho}-p_2.q\, p_1^{\rho }\right)}{p_1.q \, p_2.q} - \frac{\gamma^{\rho}_L\,t^A \otimes \cancel{q}_L \left(p_1.q-p_2.q\right)
   \left(m_d^2+m_s^2\right)}{p_1.q \, p_2.q}\nn\\
   && +\frac{\cancel{q}_L\,t^A \otimes \gamma^{\rho}_L \left(p_1.q-p_2.q\right)
   \left(m_d^2+m_s^2\right)}{p_1.q\, p_2.q}\,,\nn\\\\
 \langle O_{7} \rangle_{4q1g}  &=&2\, R \, t^A \otimes R\, m_d^2 \left(\frac{p_2^{\rho }}{p_2.q}-\frac{p_1^{\rho }}{p_1.q}\right) +q^{\nu}\,\sigma^{\nu\,\rho}_R \, t^A \otimes R\,
   m_d^2 \left(\frac{1}{p_2.q}+\frac{1}{p_1.q}\right)\nn\\
   && +q^{\nu}\, \sigma^{\nu\,\rho}_L \, t^A \otimes L \,m_s^2
   \left(\frac{1}{p_2.q}+\frac{1}{p_1.q}\right) +2\, L\,t^A \otimes L \,m_s^2 \left(\frac{p_2^{\rho
   }}{p_2.q}-\frac{p_1^{\rho }}{p_1.q}\right)\,,\nn\\\\
 \langle O_{8} \rangle_{4q1g}  &=& \frac{1}{2} q^{\nu}\, \sigma^{\nu\,\rho}_R \, t^A \otimes L\, m_d\, m_s \left(\frac{1}{p_2.q}+\frac{1}{p_1.q}\right) +L \, t^A \otimes R\, m_d\, m_s \left(\frac{p_2^{\rho }}{p_2.q}-\frac{p_1^{\rho }}{p_1.q}\right) \nn\\
&& +\frac{1}{2} q^{\nu}\,\sigma^{\nu\,\rho}_L \, t^A \otimes R\, m_d\, m_s \left(\frac{1}{p_2.q}+\frac{1}{p_1.q}\right)+R\, t^A \otimes L\, m_d\, m_s \left(\frac{p_2^{\rho}}{p_2.q}-\frac{p_1^{\rho }}{p_1.q}\right)\,.
\eea
\\
We finally report the full amplitude for the four quarks and a gluon 
\be
\mathcal{A}^{full}_{4q1g}=-i\,\frac{G_F^2}{16\, \pi^2} \, \mathrm{Im}(\lambda_c^2) \,\hat{\mathcal{A}}^{full}_{4q1g}\, ,
\ee
where
\bea
\hat{\mathcal{A}}^{full}_{4q1g}=&& \alpha _1 \left(\frac{2\, R \, t^A \otimes \cancel{q}_L\, m_d\, p_1^{\rho }}{p_1.q}-\frac{2\, \cancel{p_3}_L\,t^A \otimes \cancel{q}_L\, p_1^{\rho}}{p_1.q}+\cancel{q}_L \, t^A \otimes \cancel{p_1}_L \left(\frac{1}{p_1.q}-\frac{1}{p_2.q}\right) p_3^{\rho }
\right.  \nn\\
&& \left. +\cancel{q}_L\, t^A \otimes L  \left(\frac{1}{p_2.q}-\frac{1}{p_1.q}\right) m_s\, p_3^{\rho }-\frac{\cancel{q}_L\,t^A \otimes \cancel{q}_L\, p_3^{\rho }}{p_1.q}+2\, \cancel{p_3}_L\,t^A \otimes \gamma^{\rho}_L
\right.\nn\\
   && \left. +\gamma^{\rho}_L\,t^A \otimes \cancel{p_1}_L \left(\frac{1}{p_2.q}-\frac{1}{p_1.q}\right) p_3.q-\frac{i\, \gamma^{\alpha}_L \, t^A \otimes \cancel{p_1}_L \left(p_1.q+p_2.q\right)  \epsilon^{\alpha\, \nu\, \sigma\, \rho }\, p_3^{\nu}\, q^{\sigma}}{p_1.q\, p_2.q}\right.\nn\\
   &&\left. +\frac{i \,\gamma^{\alpha}_L\,t^A \otimes \cancel{q}_L \, \epsilon^{\alpha\, \nu\, \sigma\, \rho }\, p_3^{\nu}\, q^{\sigma}}{p_1.q}-2\, R\,t^A \otimes \gamma^{\rho}_L\, m_d+q^{\nu}\,\sigma^{\nu\,\rho}_R \, t^A \otimes \cancel{p_1}_L
   \left(\frac{1}{p_2.q}+\frac{1}{p_1.q}\right) m_d\right.\nn\\
   &&\left. -\frac{q^{\nu}\, \sigma^{\nu\,\rho}_R\,t^A \otimes \cancel{q}_L\, m_d}{p_1.q} +\gamma^{\rho}_L \, t^A \otimes L  \left(\frac{1}{p_1.q}-\frac{1}{p_2.q}\right) \,p_3.q\, m_s\right.\nn\\
   &&\left.+\frac{1}{2} \cancel{q}_L\,t^A \otimes \gamma^{\rho}_L \left(\frac{m_d^2+m_s^2-2\, p_1.p_3+4  \,p_2.q}{p_2.q}-\frac{m_d^2+m_s^2-2\, p_1.p_3+2\, p_3.q}{p_1.q}\right)\right.\nn\\
   &&\left. +\frac{i\, \gamma^{\alpha}_L\,t^A \otimes L   \left(p_1.q+p_2.q\right)  \epsilon^{\alpha\, \nu\, \sigma\, \rho }\, p_3^{\nu}\, q^{\sigma} \,m_s}{p_1.q\, p_2.q}-q^{\nu}\, \sigma^{\nu\,\rho}_R \, t^A \otimes L \left(\frac{1}{p_2.q}+\frac{1}{p_1.q}\right) m_d\,m_s \right.\nn\\
   &&\left.+i \,\gamma^{\alpha}_L\,t^A \otimes \gamma^{\nu}_L\, \epsilon^{\alpha\, \nu\, \sigma\, \rho }\, q^{\sigma} \left(\frac{ m_d^2+m_s^2-2\, p_1.p_3-2\, p_2.q}{2 \,p_2.q}+\frac{m_d^2+m_s^2-2\, p_1.p_3+2 \,p_3.q}{2 p_1.q}\right)
   \right.\nn\\
   &&\left. +\frac{1}{2}\gamma^{\rho}_L\,t^A \otimes \cancel{q}_L \left(\frac{m_d^2+m_s^2-2\, p_1.p_3+4\, p_3.q}{p_1.q}-\frac{m_d^2+m_s^2-2\, p_1.p_3+4\, p_2.q}{p_2.q}\right)\right.\nn\\
   &&\left.+\cancel{p_3}_L \, t^A \otimes \cancel{p_1}_L \left(\frac{2\, p_1^{\rho }}{p_1.q}-\frac{2\, p_2^{\rho }}{p_2.q}\right)+2\, R\, t^A \otimes L\, m_d\, m_s \left(\frac{p_1^{\rho }}{p_1.q}-\frac{p_2^{\rho }}{p_2.q}\right)
   \right.\nn\\
   &&\left.+R \, t^A \otimes \cancel{p_1}_L \left(\frac{2 \,m_d\, p_2^{\rho }}{p_2.q}-\frac{2\, m_d\, p_1^{\rho }}{p_1.q}\right)+\cancel{p_3}_L\,t^A \otimes L \left(\frac{2\, m_s\,
   p_2^{\rho }}{p_2.q}-\frac{2\, m_s\, p_1^{\rho }}{p_1.q}\right)\right.\nn\\
   &&\left.+\gamma^{\mu}_L\,t^A \otimes \gamma^{\mu}_L
   \left(-\frac{\left(m_d^2+m_s^2-2\, p_1.p_3+2 \,p_3.q\right) p_1^{\rho
   }}{p_1.q}+\frac{\left(m_d^2+m_s^2-2\, p_1.p_3\right) p_2^{\rho }}{p_2.q}+2\, p_3^{\rho
   }\right)\right)\nn\\
   && -2\, i\, \gamma^{\alpha}_L\,t^A \otimes \gamma^{\nu}_L \,\epsilon^{\alpha\, \nu\, \sigma\, \rho }\, q^{\sigma}\,,
\eea
with $\alpha_1=-\frac{5}{9}$.


\end{document}